\documentclass[%
 reprint,
 amsmath,amssymb,
 aps,
]{revtex4}

\usepackage{graphicx}% Include figure files
\usepackage{dcolumn}% Align table columns on decimal point
\usepackage{bm}% bold math
\usepackage{amsmath}
\usepackage{amssymb}
\usepackage{subfigure}
\usepackage{float}
\usepackage{tikz}
\usetikzlibrary{arrows}
\usepackage{amsthm}
\usepackage[colorlinks,linkcolor=black,anchorcolor=black,citecolor=green,CJKbookmarks=True]{hyperref}
\usepackage{titletoc}
\usepackage{graphicx}
\usepackage{epstopdf}
\usepackage[margin=10pt,font=footnotesize ,labelfont=bf, labelsep= period, width=0pt]{caption}
\usepackage{enumerate}
\usepackage[titletoc]{appendix}

\begin{document}

\preprint{APS/123-QED}

\title{W entropy in hard-core system}% Force line breaks with \\

\author{Putuo Guo}
\author{Yang Yu}%
\email{yuyang@nju.edu.cn}
\affiliation{
National Laboratory of Solid State Microstructures, School of Physics, Nanjing University, Nanjing 230093, China
}

\date{\today}% It is always \today, today,

\begin{abstract}
  As predicted by the second law of thermodynamics, the increase of entropy is irreversible in time. However, in quantum mechanics the evolution of quantum states is symmetrical about time-reversal, resulting a contradiction between thermodynamic entropy and quantum entropy. We study the W entropy, which is calculated from the probability distribution of the wave function on Wannier basis, in hard-core boson system. We find that W entropy and F entropy, which is calculated from the probability distribution of the wave function on Fock basis, satisfy an approximately linear relationship and have the same trend. Then, we investigate the evolution of W entropy for various parameters. We calculate the regression period of W entropy and find its dependence on the lattice scale. Our results show that the second law of thermodynamics is not completely valid in quantum mechanics. The behaviour of W entropy obeys the second law of thermodynamics, only when the system scale is large enough.
% \begin{description}
% Secondary publications and information retrieval purposes.
% You may use the \texttt{description} environment to structure your abstract;
% use the optional argument of the \verb+\item+ command to give the category of each item. 
% \end{description}
\end{abstract}

\maketitle

\section{\label{sec:level1}Introduction}

The second law of thermodynamics states ``the entropy of an isolated system does not decrease spontaneously'', predicting that the entropy always increases with time. If we write the entropy as a function of time $S(t)$, then according to the second law, for the time series $t=t_1,t_2,\ldots,t_n$, $S(t_n)\geqslant S(t_{n -1})\geqslant\ldots\geqslant S(t_2)\geqslant S(t_1)$ should be satisfied. When we reverse the time series, $t=t_n,\ldots,t_2,t_1$,  $S(t_1)\geqslant S(t_{2})\geqslant\ldots\geqslant S(t_{n-1})\geqslant S(t_n)$ is obviously not true unless $S(t)$ is a constant. Therefore, the second law of thermodynamics is not symmetrical about time-reversal. However, in dynamics both Newton's law and Schrodinger's equation satisfy the time-reversal symmetry. The classic Hamiltonian canonical equations are,
\begin{align}
\frac{\partial H}{\partial q}=-\frac{dp}{dt}\\
\frac{\partial H}{\partial p}=\frac{dq}{dt}\;.
\end{align}
The Schrodinger equation is,
\begin{align}
  H\psi=i\hbar\frac{d}{d t}\psi\;.
\end{align}
When we substitute $t=-t$, the above two equations still hold. In 1874, Thomson proposed that the second law of thermodynamics does not satisfy the time-reversal symmetry \cite{Loschmidtparadox}. In 1876, Loschmidt pointed out that we can not get the irreversible process from dynamic equations only in mathematical form. This is called Loschmidt's paradox. It is generally believed that the second law of thermodynamics is usually a description of multi-particle or macrosystems, and does not hold under a single particle or a few particles. However, in 1890, Poincare presented the Poincare regression theorem \cite{recurrencetheorem}, which states that any dynamic systems will, after a sufficiently long but finite time, return to a state arbitrarily close to their initial state. In order to study whether the second law of thermodynamics is also valid in quantum mechanics, it is necessary to define a quantum entropy corresponding to Boltzmann's statistical entropy. According to the uncertainty principle, the position and momentum cannot be determined at the same time, thus each point in classical phase space loses its meaning. In 1929, von Neumann proposed a method of establishing a quantum phase space by constructing commensurable macro-position operator $Q$ and macro-momentum operator $P$. He proved the ergodic theorem of quantum states and the quantum H theorem \cite{EHTV}. $Q$ and $P$ satisfy the following relations.
\begin{align}
  \left[Q,P\right]=0\\
  Q\sim q,\quad P\sim p
\end{align}
where $q$ is position operator and $p$ is momentum operator, which satisfy $[q,p]=i\hbar$. Mathematically, finding $Q$ and $P$ are equivalent to finding a set of wave functions $\{w_j\}$. Here, $w_j$ is local in both position space and momentum space. The macro operators can be expressed by this set of wave functions as follow.
\begin{align}
  P&=\sum_j\left|w_j\right>\left<w_j\right|p\left|w_j\right>\left<w_j\right|,\\
  Q&=\sum_j\left|w_j\right>\left<w_j\right|q\left|w_j\right>\left<w_j\right|
\end{align}
Then we can calculate the $i$th level difference between macro and micro operators using 
\begin{align}
  \Delta^{(i)}p_j&=\left<w_j\left|\left(p-\left<p\right>_j\right)^i\right|w_j\right>^{1/i},
  \label{eq:Dp}
  \\
  \Delta^{(i)}q_j&=\left<w_j\left|\left(q-\left<q\right>_j\right)^i\right|w_j\right>^{1/i}
  \label{eq:Dq}
\end{align}
When $i\geqslant 2$, the above two differences should be sufficiently small.
\par
For one-dimensional system, $q=x,\quad p=\hbar k=-i\hbar\partial_x$, von Neumann proposed to perform Schmidt orthogonality on a set of Gauss wave packets with a width of $\zeta$. The result is $\{w_j\}$.
\begin{align}
  g_{j_x,j_k}=\exp\left[-\frac{(x-j_xx_0)^2}{4\zeta^2}+ij_kk_0x\right]
\end{align}
This method, which is called ``cumbersome'' by von Neumann, suffers from two major drawbacks. The first is that it cannot be tested by numerical calculation because of the large amount of calculation, and the result is sensitive to the order of orthogonalization. The second point is that von Neumann believes, the existence of $Q$ and $P$ depends on the fact that the momentum and position can be measured simultaneously in macroscopic measurements. Finally, the function set does not satisfy spatial translation symmetry. 
\par
In 2015, Han and Wu established a quantum phase space \cite{qhtheorem,qphasespace} by constructing a set of Wannier functions that are local in both momentum and position. The pure quantum state can be mapped to quantum phase space unitarily to get its probability distribution, which is used to define its W entropy. They proved the inequality about W entropy's long-term fluctuation behaviour, and made numerical calculation in case of single-particle \cite{qhtheorem}.
\par
\begin{figure}[H]
\includegraphics[width=8cm,height=6cm]{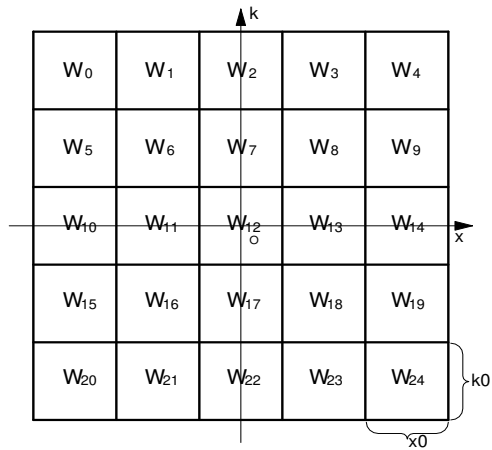}
\centering
\caption{Schematic diagram of two-dimensional quantum phase space. The length and width of each cell are $x_0$ and $k_0$.  Each cell is assigned a Wannier function $\left|w_j\right>$. For $\left|\psi\right>$, its projection on cell $j$ is $\left<w_j\big|\psi\right>$.}
\label{fig:wlattice}
\end{figure}

In quantum mechanics, position and momentum satisfy $\Delta x\Delta p\geqslant\hbar/2$. Defining each point with a precise $x$ and $p$ in the phase space no longer makes sense. We divide the phase space into a series of cells with $x_0$ as length and $p_0$ as width, where $x_0p_0=h$, and $h$ is Planck constant. The volume of each phase cell is Planck constant, also called Planck cell. For the convenience of calculation, we replace momentum $p$ with wavenumber $k$, $p=\hbar k$, so that $x_0$ and $k_0$ satisfy the following relationship.
\begin{align}
  x_0k_0=x_0p_0/\hbar=2\pi\quad ,
\end{align}
where $\hbar=h/2\pi$. As shown in Fig. \ref{fig:wlattice}, each phase cell corresponds to a Wannier function at local position. The wave function is mapped to quantum phase space by its inner product with the Wannier function. For quantum state $\psi$, the probability on phase cell $j$ is
\begin{align}
  p_j=\left|\left<w_j\big|\psi\right>\right|^2
  \label{Probability}
\end{align}
The W entropy is defined as
\begin{align}
  S_w=-\sum_jp_j\ln p_j\quad,
  \label{Def_of_Sw}
\end{align}
where $p_j$ is defined by the Eq. (\ref{Probability}). For mixed state, the definition can be generalized as follows. The density matrix of $\psi$ is $\rho=\left|\psi\big>\big<\psi\right|$, and we can project it on Wannier basis to get
\begin{align}
  \rho_{ij}=\left<w_i\big|\psi\right>\left<\psi\big|w_j\right>\quad .
  \label{rho_on_w}
\end{align}
When $\psi$ is pure state, $p_j=\rho_{jj}$. The W entropy of mixed state can be defined as
\begin{align}
  S_w=-\sum_j\rho_{jj}\ln\rho_{jj}\quad .
  \label{mix_state_Sw}
\end{align}
\par
Following the above discussion, quantum entropy will remain constant or oscillate for a simple system (such as a harmonic oscillator). For complex systems, a mainstream view is that when we take $t=-t$, the quantum entropy will decrease and return to the initial value. However, the decrease and increase can be regarded as ``accidental'' fluctuations in a large period, and the probability of occurrence of the entropy decrease event is very low, thus it will not be observed. This explains the contradiction between Boltzmann entropy and quantum entropy. However, in quantum mechanics, except for quantum measurement, the evolution of a system is deterministic. It seems a bit far-fetched to explain the second law of thermodynamics through ``accidental'' fluctuations in deterministic theory. Another argument is that such large fluctuations have periodicity, but the period is so long that it cannot be measured and calculated. We believe that whether it fits probabilistic explanation or periodic explanation, the fundamental reason is the complexity of system. As the complexity gradually increases, the return period of quantum entropy will become longer and longer, approach to Boltzmann entropy eventually.
\par
To study the relationship between the regression period and the complexity, we calculate the W entropy of hard-core boson system, mainly based on four reasons as below. Firstly, W entropy is the equivalent of Boltzmann entropy in quantum mechanics. Secondly, the system must be able to increase the scale gradually. An overly complicated and simple system cannot find the intermediate state of periodic changes. Thirdly, the dimension of Hilbert space is not too high. Fourthly, the system is isolated. We calculated the W entropy in different shape systems and obtained an approximately linear relationship between $S_w$ and $S_f$ (entropy under the Fock basis) for the first time. It shows that $S_w$ and $S_f$ have the same trend. The curve that the regression period increased by lattice number is plotted. It is also found that the regression period will increase drastically as the number of particles increases. Finally, the physical explanation of $S_w(t)$ is given. W entropy does not destroy the time-reversal symmetry in the hard-core boson system. It has a certain regression period, but as the system scales up, its period becomes very long, and $S_w(t)$ in one period approaches to $A'(1-\exp(-\omega t))+b$. The second law of thermodynamics is not entirely valid in quantum mechanics. The W entropy obeys the second law of thermodynamics, only when the system scale is large.

\section{Hard-core boson system}
\begin{figure}[H]
	\centering
	\includegraphics[scale=0.40]{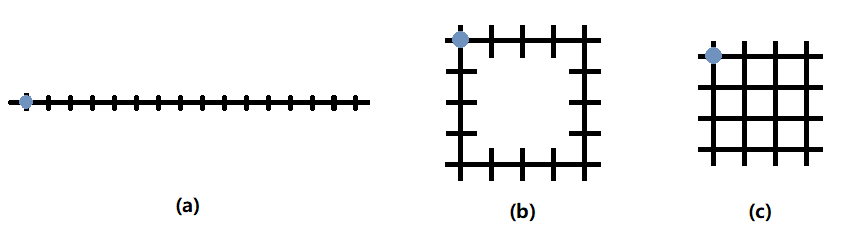}
	\caption{Different shape lattices. (a) chain; (b) ring shaped; (c) square}
	\label{fig:scc_1}
\end{figure}

The Hamiltonian of Bose-Hubbard model without chemical potential under Fock basis is
\begin{align}
	H=-J\sum_{\left<i,j\right>}b_i^\dag b_j+\frac{V}{2}\sum_in_i(n_i-1).
  \label{eq:H_Hubbard}
\end{align}
where $<i,j>$ indicates that the sums run over all nearest-neighbor pairs of sites, $J$ is the hopping parameter, and $V$ is the on-site interaction.
The boson creaton ($b_i^\dag$) and annihilation ($b_j$) operators commute on different sites.
\begin{align}
  [b_i,b_j^\dag]&=[b_i,b_j]=[b_i^\dag,b_j^\dag]=0\quad\forall i\; and\; i\neq j\\
  \{b_i,b_i^\dag\}&=1\quad and \quad (b_i)^2=(b_i^\dag)^2=0\quad \forall i\\
  n_i&=b_i^\dag b_i
  \label{eq:bibj}
\end{align}
Here $n_i=b_i^\dag b_i$ is the density operator. 
%Its Hamiltonian has only two terms. One is the momentum term, which allows particles to a tunnel (``jump'') between two sites. The other is the potential term, which describes the interaction between particles on the lattice.
The hard-core boson system is the result of the Bose-Hubbard model taking the hard-core boson approximation. % The Hamiltonian 
For simplicity, we take the hard-core limit (the energy level of each position on the lattice can only be 0 or 1), $V\to\infty$, let $J$ fixed, and add the interaction term $U$. The Hamiltonian of the model is given by
\begin{align}
  H=-J\sum_{\left<i,j\right>}b_i^\dag b_j+U\sum_{\left<i,j\right>}n_in_j
  \label{eq:H_HardCore}
\end{align}
Here, $i$ and $j$ represent the site on the lattice. Each site represents a harmonic oscillator. There are many shapes of lattices. Fig. \ref{fig:scc_1}(c) shows one of the 4$\times$4 lattices.
The first term of the Hamiltonian indicates the kinetic energy of the particle from position $i$ to $j$. The second term represents the interaction between the particle at position $i$ and position $j$. $<i,j>$ indicates that the sums run over all nearest-neighbour pairs of sites. For Fig. \ref{fig:scc_1}(c), if $i=1$ then $j=2,5$. Hard-core means that only one boson can exist in the same position, so the energy level of harmonic oscillator is 0 or 1. 

%%%%%%%%%%%%%%%%%%%%%%%%%%%%%%%%%%%%%%%%%%%%%%%%%%%%%%%%%%%%%%%%%%%

\section{W entropy changes over time}
For convenience of description, we define F entropy as the entropy calculated by the probability distribution of quantum state under Fock basis in hard-core boson system. For $\psi=\sum_m \lambda_m\left|m\right>$, F entropy is defined as
\begin{align}
S_f=-\sum_i p_i\ln p_i,
\end{align}
where $p_i=|\lambda_i|^2$. Define $n$ to be the number of lattice sites, and $N$ to be the number of particles. We found that in hard-core boson systems F entropy and W entropy satisfy the linear relationship approximately as bellow.
\begin{align}
S_w\approx kS_f+b 
\label{eq:liner}
\end{align}
where $k$ and $b$ are real constants, and $b>0$. To study how particle number, system scale and lattice shape affect the W entropy, we calculated the following four cases:
\begin{enumerate}
\item fixed particle number and shape of lattice, changed lattice scale.
\item fixed scale and shape of lattice, changed the particle number.
\item fixed scale and shape of lattice and particle number,  the initial position is changed.
\item fixed the scale of lattice and particle number,  the shape becomes chain, ring, two-dimensional.
\end{enumerate}

We will discuss these four cases separately below.
\par
\begin{figure}[H]
\centering
\includegraphics[scale=0.7]{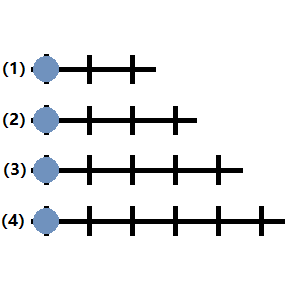}
\caption{The diagram shows lattices of n from 3 to 6, fixed $N=1$ and initially on the far left. The labels from (1) to (4) correspond to the different curves in Fig. \ref{fig:Sw_t_n} and \ref{fig:n_wf}.}
\label{fig:diff_n}
\end{figure}
\begin{figure}[H]
\centering
\includegraphics[scale=0.4]{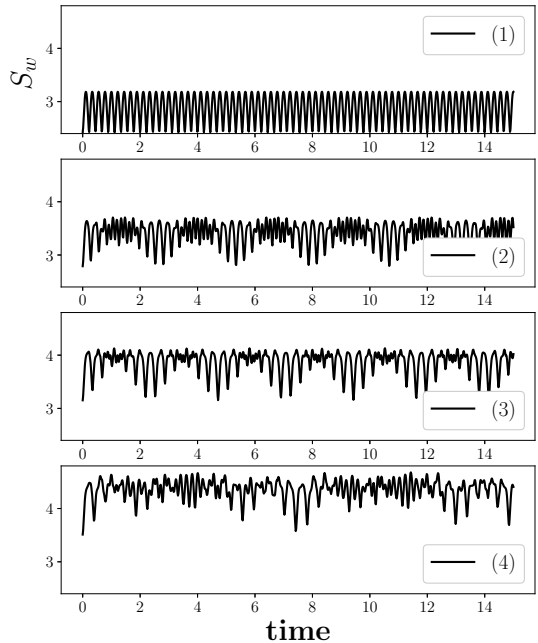}
\caption{The W entropy changes with time. Label (1) to (4) corresponds to the systems in Fig. \ref{fig:diff_n}. Both minimum and maximum values of W entropy increase and the regression period becomes longer.}
\label{fig:Sw_t_n}
\end{figure}
\begin{figure}[H]
\centering
\includegraphics[scale=0.35]{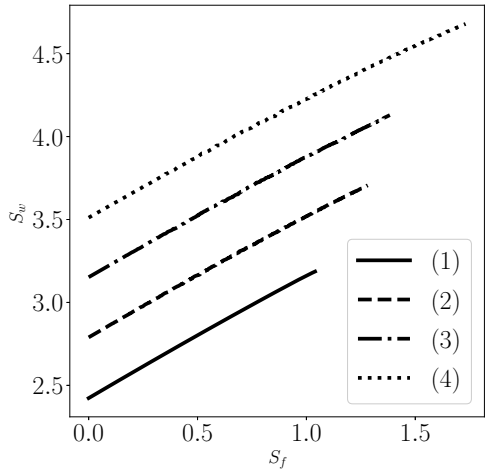}
\caption{The linear relationship (Eq. (\ref{eq:liner})) between W entropy and F entropy. Label (1) to (4) corresponds to the systems in Fig. \ref{fig:diff_n}. The slopes of lines from (1) to (4) decrease gradually as shown in Fig. \ref{fig:n_k}. From (1) to (4), their intercepts (b) increase linearly. The slope is decreasing, although this decrease is not significant there. When we make Fig. \ref{fig:n_k}, it does drop a bit. So there is a linear relationship between W entropy and F entropy, although the linear coefficient will be different for different n.}
\label{fig:n_wf}
\end{figure}

First, we calculated $S_w(t)$ and $S_f(t)$ with different $n$ under fixed $N=1$. The shape of the lattices is shown in Fig. \ref{fig:diff_n}. $S_w(t)$ is shown in Fig. \ref{fig:Sw_t_n}. The linear relationship between W and F entropy is shown in Fig. \ref{fig:n_wf} We find that as $n$ increases, $k$ decreases, while $b$ increases linearly. The initial rising speed of W entropy keeps unchanged. The maximum value becomes larger, and the regression period becomes longer. 

\begin{figure}[H]
\centering
\includegraphics[scale=0.35]{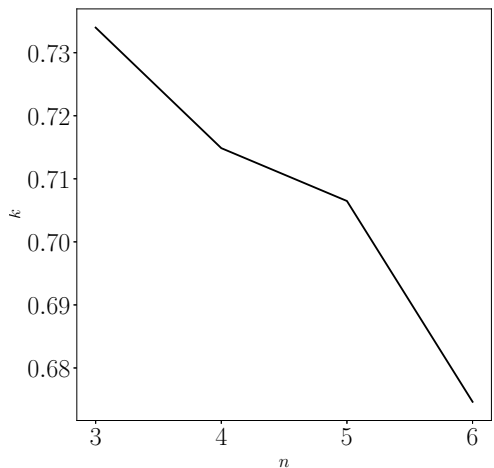}
\caption{The Slope ($k$) of lines in Fig. \ref{fig:n_wf} changes with $n$. k decreases slightly when $n$ increases from 3 to 6. }
\label{fig:n_k}
\end{figure}

%\begin{figure}[H]
%\centering
%\includegraphics[scale=0.35]{n_b.eps}
%\caption{The intercept ($b$) of lines in Fig. \ref{fig:n_wf} changes with $n$. For each curve in Fig \ref{fig:Sw_t_n}, $S_w(t=0)$ also equals to $b$. $b$ increases linearly with $n$.}
%\end{figure}

\par
\begin{figure}[H]
\centering
\includegraphics[scale=0.7]{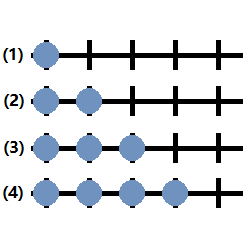}
\caption{The particle number $N$ from $1$ to $4$, fixed $n=5$. The labels from (1) to (4) correspond to the different curves in Fig. \ref{fig:1234_5} and \ref{fig:0N_wf}. The blue circle on the left represents the initial particles. Although the systems (1) and (4), (2) and (3) are dynamically equivalent, as the number of particles N increases, the W entropy of the system will increase.}
\label{fig:diff_0N}
\end{figure}
\begin{figure}[H]
\centering
\includegraphics[scale=0.4]{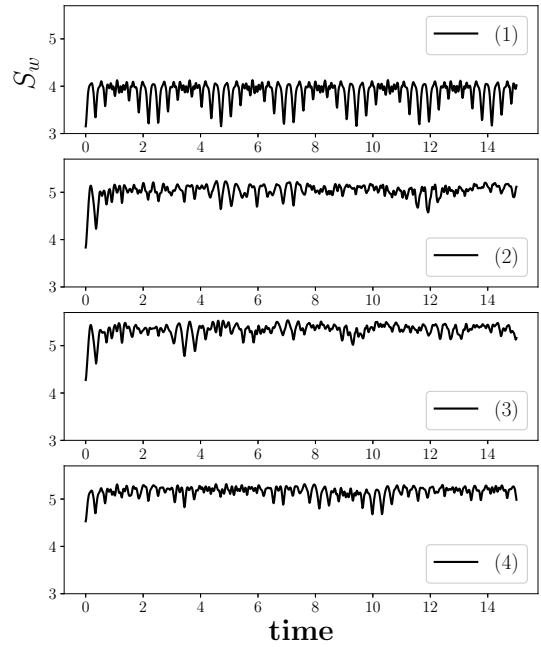}
\caption{W entropy of systems in Fig. \ref{fig:diff_0N}. As the number of particles N increases, the W entropy of initial state is getting larger.  F entropy and W entropy can maintain a linear relationship.}
\label{fig:1234_5}
\end{figure}
\begin{figure}[H]
\centering
\includegraphics[scale=0.38]{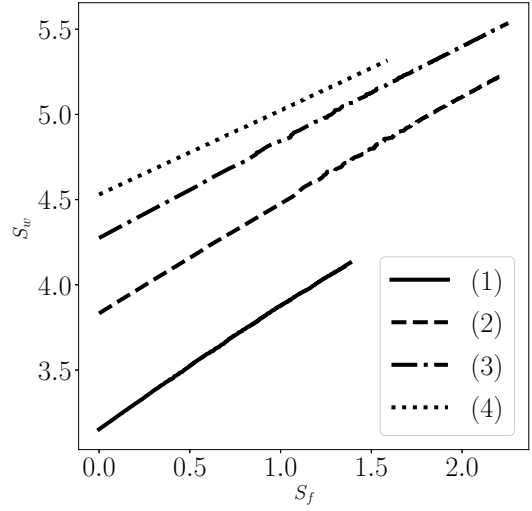}
\caption{The linear relationship (Eq. (\ref{eq:liner})) between W entropy and F entropy. Label (1) to (4) corresponds to the systems in Fig. \ref{fig:diff_0N}. The slope ($k$) from (1) to (4) decreases, while the intercept ($b$) increases, but the increase rate slows down.}
\label{fig:0N_wf}
\end{figure}

Secondly, in Fig. \ref{fig:diff_0N}, we fixed $n=5$, increased $N$ from 1 to 4. The evolution result is shown in Fig. \ref{fig:1234_5} and the linear relationship is in Fig. \ref{fig:0N_wf}. When $N$ increases, $k$ decreases and $b$ increases. The maximum value of W entropy changes. When $0<N<n/2$, the periodicity of the entropy decreases, and the fluctuations become smaller. Compared with increasing $n$, increasing $N$ has more influence on the regression period.
\par
From the definition of F entropy, we know that $\left|0\right>\otimes\left|0\right>$ and $\left|0\right>\otimes\left|1\right>$ have the same F entropy . However, since the W entropy of $\left|1\right>$ is larger than that of $\left|0\right>$, although (1) and (4) ((2) and (3)) in Fig. \ref{fig:diff_0N} are dynamically equivalent (particle-hole transformation), but their W entropy is different.

%\begin{figure}[H]
%\centering
%\includegraphics[scale=0.38]{0N_b.eps}
%\caption{Intercept ($b$) of lines in Fig. \ref{fig:0N_wf} changes with $N$. When $N$ increase from 1 to 4, b increases.}
%\label{fig:b_0N}
%\end{figure}
\par
\begin{figure}[H]
\centering
\includegraphics[scale=0.7]{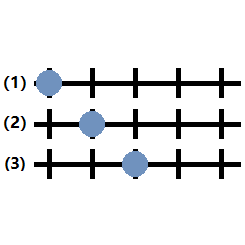}
\caption{This diagram shows the lattices of n from 3 to 6, fixed N=1 and n=5. The labels from (1) to (3) correspond to the different curves in Fig. \ref{fig:position_1} and \ref{fig:p1_wf}. The blue circle on the left represents the initial state particles. The choice of location does not affect the linear relationship between W entropy and F entropy.}
\label{fig:diff_p1}
\end{figure}
\begin{figure}[H]
\centering
\includegraphics[scale=0.4]{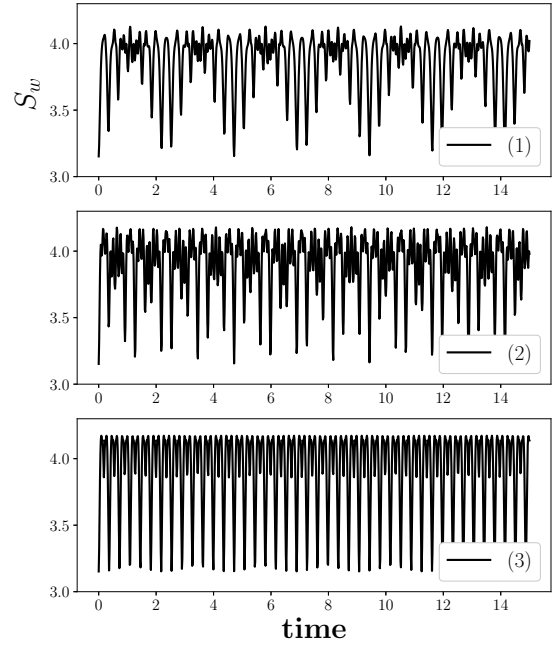}
\caption{This figure shows the W entropy changes by time. Label (1) to (3) corresponds to the systems in Fig. \ref{fig:diff_p1}. They have the same minimum and maximum values, and similar regression periods.}
\label{fig:position_1}
\end{figure}
\begin{figure}[H]
\centering
\includegraphics[scale=0.38]{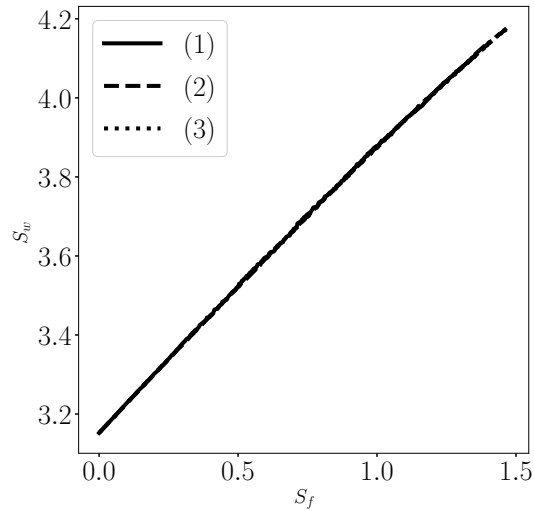}
\caption{The linear relationship between W entropy and F entropy. Label (1) to (3) corresponds to the systems in Fig. \ref{fig:diff_p1}. The lines overlap each other. It shows that the different initial positions of the particles do not affect the linear relationship. As long as the number of grid points n and the number of particles N are unchanged, the linear coefficients are the same.}
\label{fig:p1_wf}
\end{figure}

Next, we discuss the influence of particle location on $S_w(t)$. As shown in Fig. \ref{fig:diff_p1}, the five lattice sites system with one particle, we just change the position. The evolution result is shown in Fig. \ref{fig:position_1}. The linear relationship keeps unchanged in Fig. \ref{fig:p1_wf}.
$k$ and $b$ are constant when changing the particle's initial position because the initial position does not relate to $N$ and the dimensions of Hibert space. Their regression period is also close. Besides, the result of the two particles is the same as that of single.

\begin{figure}[H]
\centering
\includegraphics[scale=0.4]{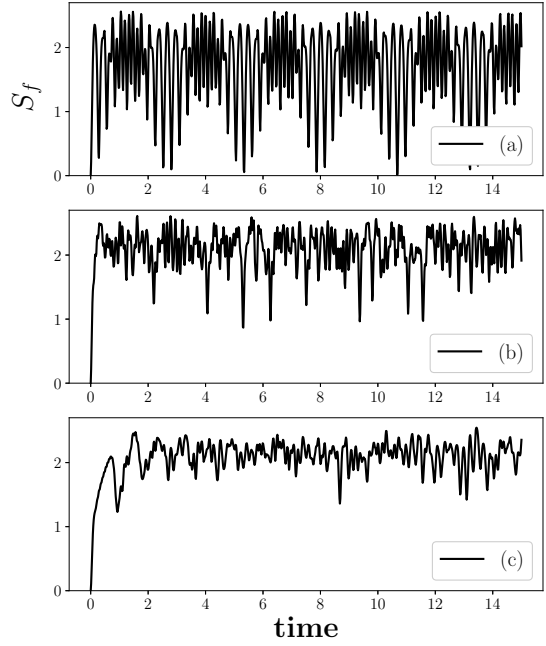}
\caption{The effect of lattice shape on the change of W entropy. Fig. (a), (b), and (c) respectively show the F entropy of the corresponding label lattice in the Fig. \ref{fig:scc_1} as a function of time. Because F entropy and W entropy satisfy an approximately linear relationship, their trend and regression period are equal to W entropy. It can be seen from (a), (b) and (c) that as the number of connection sites in the lattice increases, the regression period of W entropy decreases.}
\label{fig:scct}
\end{figure}

Finally, we discuss the influence of different connection modes on $S_w(t)$. Different connection modes do not affect $N$ and $n$ too. It does not change the values of $k$ and $b$.
We calculated the F entropy of chain, ring shaped and square under 16 sites in Fig. \ref{fig:scc_1}. Since F entropy and W entropy have a linear relationship, their periods and fluctuations are the same. The calculation result is shown in Fig. \ref{fig:scct}. It can be seen that as the number of connection sites increases, the regression period decreases.

%\begin{figure}[H]
%\centering
%\includegraphics[scale=0.38]{p1_label_b.eps}
%\caption{Intercept ($b$) of lines in Fig. \ref{fig:p1_wf}. $b$ keeps constant with the change of initial position of the particle.}
%\end{figure}

\par

Both Fock basis and Wannier basis are complete and orthogonal. Because the dimensionality of Wannier basis is higher, there will be redundancy in the representation process, and $S_w$ will have at least one more constant than $S_f$. If a Fock basis is taken, its $S_f=0$, $b\approx S_w$. $b$ is the Fock basis's average value of W entropy.
\begin{align}
b=\overline S_w(fock)
\end{align}
The F entropy of $\left|\psi\right>=\frac{1}{\sqrt 2}\left(\left|0\right>+\left|1\right>\right)$ is $S_f= \ln 2$. Due to the increase in the Hilbert space dimension, the probability of $\psi$ on the Wannier function set is equivalent to a certain degree of ``amortization'' of the probability distribution, and $k$ describes the average proportion of ``amortization''. It shows that W entropy and F entropy have the same changing trend over time, and their first derivative with time has the following relationship.$\frac{dS_w}{dt}\approx k\frac{dS_f}{dt}$. However, whether the linear relationship holds in any two orthogonal normalized bases remains to be further studied. For example, when the selected basis is close to the energy representation, $k\to 0$, each basis has a different energy. When approaching to the particle number representation, $k$ reaches its maximum value, but the energy of each basis is the same.
\par
\begin{figure}[H]
\centering

\includegraphics[scale=0.4]{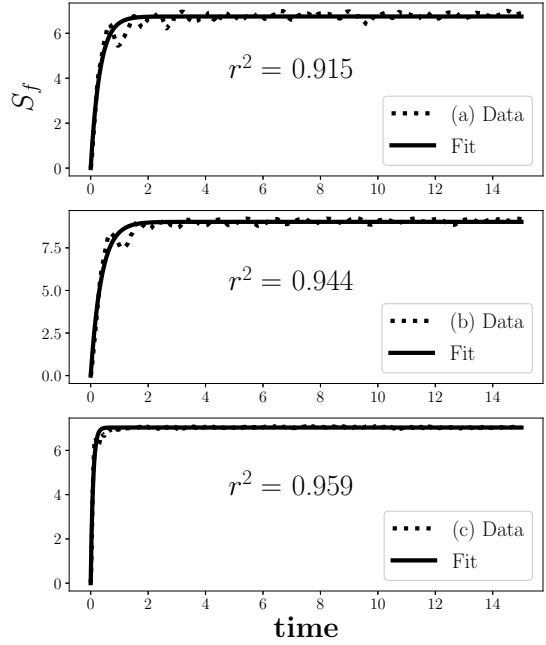}
\centering
\caption{(a) F entropy of $n=16,\; N=4$ chain; (b) F entropy of $n=18,\; N=6$ chain; (c) F entropy of $n=16,\; N=4$ square. The solid line is the result of fitting by the least square method of Eq. (\ref{eq:Sf_fit}), and the dotted line is the data obtained by the numerical solution of the Schrodinger equation. When the scale of the system becomes more extensive, the curve and Eq. (\ref{eq:Sf_fit}) getting closer and closer, the goodness of fit $r^2$ gradually approaches to 1.
}
\label{fig:fit}
\end{figure}

Secondly, when the periodicity weakens, and the fluctuation decreases, F entropy in a period with time approaches to a simple function as bellow.
\begin{align}
S_f=A(1-e^{-\omega t})
\label{eq:Sf_fit}
\end{align}
F entropy of (a) $n=16, N=4$ chain, (b) $n=18, N=6$ chain, (c) $n=16, N=4$ square is plotted as dotted line in Fig. \ref{fig:fit}. We use the least square method to fit them with Eq. \ref{eq:Sf_fit}. Setting $Fit$ to be the F entropy that calculated from Eq. \ref{eq:Sf_fit} (solid line in Fig. \ref{fig:fit}) and $Data$ to be the F entropy (numerical result of hard-core boson system), the total sum of squares ($SST$) and error sum of squares ($SSE$) could be expressed as 
\begin{align}
SST=\sum_i{(Data_i-\overline{Data})^2}\\
SSE=\sum_i{(Data_i-Fit_i)^2}\; .
  \label{eq:sset}
\end{align}
Then, the goodness of fit ($r^2$) is calculated by
\begin{align}
  r^2=1-\frac{SSE}{SST}\; .
  \label{eq:rsquare}
\end{align}
In Fig. \ref{fig:fit}, it can be seen that as the scale of the system becomes larger, the solid line (fitting result) and dotted line (numerical result) are getting closer and closer, the goodness of fit $r^2$ gradually approaches to 1.

F entropy is expressed as 
\begin{align}
\lim_{n,N\to\infty}S_f(t)\approx A(1-\exp(\omega t))
\label{eq:Sf_limit}
\end{align}
Substituting into the linear relationship (Eq. \ref{eq:liner}), the approximate limit of W entropy in one period can be expressed as 
\begin{align}
\lim_{n,N\to\infty}S_w(t)\approx A'(1-\exp(\omega t))+b\quad ,
\end{align}
where $A'=kA$. 
\par
\begin{figure}[H]
  \centering
  \includegraphics[scale=0.3]{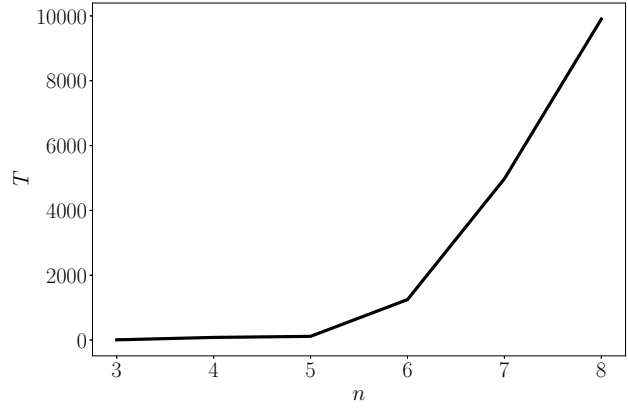}
  \caption{The period varies with the number of sites. As the number of sites increases, the regression cycle will become longer and longer. At the same time, the more sites, the faster the growth.}
  \label{fig:T_n}
\end{figure}

According to the regression theorem, when the time is sufficiently long, the evolution of the hard-core boson system can be infinitely close to the initial state. We use W entropy to describe the hard-core boson system. Suppose that 
\begin{align}
\epsilon=S_w(T)-S_w(0)\quad ,
\end{align}
where $T$ is regression period. If $\epsilon<0.2$, we determine that W entropy returns. We calculate the regression periods by n from 3 to 8, shown in Fig. \ref{fig:T_n}. The increasing speed of $T(N)$ is much greater than that of $T(n)$, as $T(N=2)>10000$.

%%%%%%%%%%%%%%%%%%%%%%%%%%%%%%%%%%%%%%%%%%%%%%%%%%%%%

\section{Numerical method}
First, we can express $\left|0\right>$ and $\left|1\right>$ as the summation of Wannier functions as bellow.
\begin{align}
\left|0\right>=\sum_iC_{0i}w_i\\
\left|1\right>=\sum_iC_{1i}w_i
\end{align}
Second, the basis of Fock state can be expressed as Wannier functions. In the case of two nodes,  we write the Fock basis as 
\begin{align}
\left|m\right>\otimes\left|n\right>=\sum_iC_{mi}w_i\otimes\sum_jC_{nj}w_j,
\end{align}
 where $m,n=0,1$. Then we can transform the quantum state from Fock basis to Wannier basis easily. 
 \begin{align}
\left|\psi\right>&=\sum_{m,n=0,1}C'_{mn}\left|m\right>\otimes\left|n\right>\\
&=\sum_{m,n,i,j}C'_{mn}C_{mi}C_{nj}w_i\otimes w_j.
\end{align}
The probability distribution in quantum phase space is $p_{ij}=\sum_{m,n}\left|C'_{mn}C_{mi}C_{nj}\right|^2$. The W entropy is calculated by $S_w=-\sum_{ij}p_{ij}\ln p_{ij}$.  This method can be generalized to the case of $n$ nodes quickly.

\section{Conclusion}
In summary, we have calculated the W entropy over time in hard-core boson system and found that W entropy and F entropy satisfy an approximately linear relationship. The two linear coefficients respectively describe the W entropy of Fock basis and the coefficient of wave function contraction in basis transformation. This shows that W entropy and F entropy have the same trend over time. By calculating the systems with different $n$, $N$ and connection nodes, we found that the regression period $T$ will increase as $n$ and $N$ increase, and is negatively correlated with the number of connections between lattice sites. $T(n)$ shows that the W entropy's regression period changes with the number of lattice sites. The position of the initial particles has no influence on $T$. Finally, the physical explanation of $S_w(t)$ is given. W entropy does not destroy the time-reversal symmetry in the hard-core boson system. But as the system scales up, its period becomes very long, and approaches to $A'(1-\exp(-\omega t))+b$ in one period. The second law of thermodynamics is not completely valid in quantum mechanics. Only when the scale of system is large, the behaviour of W entropy obeys the second law of thermodynamics.

\par

Although the D entropy \cite{AnatoliPolkovnikov2010} of the energy eigenstate of the system does not change with time, the subsystems's D entropy is time-dependent. If the relationship between D entropy and W entropy of the subsystem can be found, the relationship between W entropy and the density of state can be found, and it can be connected with thermodynamic entropy directly. One of the differences between W entropy and Boltzmann entropy is that Boltzmann entropy describes the observed result of quantum state and W entropy describes the state before measured. Studying the relationship between W entropy and Boltzmann entropy is helpful to further explore the physical meaning of quantum measurement. The W entropy can be further used to analyze the quantum heat engine, which can work between the heat source and the cold source like the classical heat engine. The Hamiltonian of one of the spin Bose models \cite{heatengine} is as follows.
$$
H=\frac{\hbar\Omega}{2}\sigma_z+\hbar\omega\left(a^\dag a +\frac{1}{2}\right)+\hbar\omega x_0\sigma_z\frac{ a+a^\dag}{\sqrt 2}
$$
$\sigma_z=\left|e\right>\left<e\right|-\left|g\right>\left<g\right|$. The first term on the left is the thermal power conversion term, the second is the harmonic oscillator term, and the third is the qubit term. In addition to the spin model, there are other quantum heat engine models. Their principles are roughly the same. They both couple the cold hot bath to a qubit and then couple the qubit to the harmonic oscillator. We can try to replace the cold bath with a hard-core boson model, couple it with a quantum heat engine, and then calculate the change in W entropy and analyze the efficiency of the quantum heat engine. If the second law of thermodynamics can be broken when the system scale is small enough, the efficiency of this kind of quantum heat engine may exceed the Carnot heat engine. It is also a direction to study whether there are factors other than the second law that restrict the efficiency of the quantum heat engine. 

%\addcontentsline{toc}{section}{References}%将“参考文献加入目录中”
%\bibliographystyle{unsrt} %设置参考文献类型,unsrt,alpha ,abbrv
%\bibliography{template} %插入参考文献
\section*{Acknowledgements}
This work was partly supported by the NKRDP of China (Grant No.2016YFA0301802), NSFC (NO.61521001, and No.11890704), and the Key R\&D Program of Guangdong Province (Grant No.2018B030326001). 
\par
Special thanks to Prof. Wu Biao, Physics College of Peking University. We got much inspiration and help through several discussions, with Prof. Wu during the study of this paper.
\par
We used some functions in QuTiP \cite{qutip,qutip2} in numerical calculations. Thank the developers.
\begin{appendices}
\end{appendices}

\appendix

\bibliography{manuscript}% Produces the bibliography via BibTeX.

\end{document}